\documentclass[twocolumn,aps,prb,twoside,superscriptaddress,floatfix,nofootinbib]{revtex4-1}

\usepackage{dcolumn}
\usepackage{float}
\usepackage{bm}
\usepackage[T1]{fontenc}
\usepackage{amsmath}
\usepackage{graphicx} 
\usepackage{tabularx}
\usepackage{multirow}
\usepackage{rotating} 
\usepackage{makecell}
\usepackage{wasysym}
\usepackage{hyperref}
\usepackage{pifont}
\hypersetup{colorlinks=true, linkcolor=blue, citecolor=blue, urlcolor=blue}
\usepackage{tikz}
\usepackage[normalem]{ulem}

\makeatletter
\renewcommand\@biblabel[1]{#1.}
\makeatother

\makeatletter
\newcommand*{\citenumns}[2][]{%
  \begingroup
  \let\NAT@mbox=\mbox
  \let\@cite\NAT@citenum
  \let\NAT@space\NAT@spacechar
  \let\NAT@super@kern\relax
  \renewcommand\NAT@open{}%
  \renewcommand\NAT@close{}%
  \cite[#1]{#2}%
  \endgroup
}
\makeatother

\newcommand{\super}[1]{\ensuremath{^{\textrm{#1}}}}
\newcommand{\sub}[1]{\ensuremath{_{\textrm{#1}}}}

\begin{document}

\title{Unconventional order-disorder phase transition in improper ferroelectric hexagonal manganites}
\author{Sandra H. Skj\ae rv\o}
\email{sandra.skjaervoe@psi.ch}
\affiliation{NTNU Norwegian University of Science and Technology, Department of Materials Science and Engineering, NO-7491 Trondheim, Norway}
\altaffiliation[Current affiliation: ]{Laboratory for Mesoscopic Systems, Department of Materials, ETH Zurich, CH-8093 Zurich, Switzerland, Laboratory for Multiscale Materials Experiments, Paul Scherrer Institute, CH-5232 Villigen PSI, Switzerland}
\author{Quintin N. Meier}
\affiliation{ETH, Materials Theory, Wolfgang Pauli Str. 27, CH-8093 Z\"{u}rich, Switzerland}
\author{Mikhail Feygenson}
\affiliation{Forschungszentrum J\"ulich, JCNS, D-52425 J\"ulich, Germany}
\affiliation{Chemical and Engineering Materials Division, Oak Ridge National Laboratory, Oak Ridge, TN 37831, USA}
\author{Nicola A. Spaldin}
\affiliation{ETH, Materials Theory, Wolfgang Pauli Str. 27, CH-8093 Z\"{u}rich, Switzerland}
\author{Simon J. L. Billinge}
\affiliation{Brookhaven National Laboratory, Condensed Matter Physics and Materials Science Department, Upton, NY 11973, USA}
\affiliation{Columbia University, Department of Applied Physics and Applied Mathematics, New York, NY 10027, USA}
\author{Emil S. Bozin}
\affiliation{Brookhaven National Laboratory, Condensed Matter Physics and Materials Science Department, Upton, NY 11973, USA}
\author{Sverre M. Selbach}
\email{selbach@ntnu.no}
\affiliation{NTNU Norwegian University of Science and Technology, Department of Materials Science and Engineering, NO-7491 Trondheim, Norway}
\begin{abstract}
\label{sec:abstract}
The improper ferroelectricity in YMnO$_3$ and other related multiferroic hexagonal manganites are known to cause topologically protected ferroelectric domains that give rise to rich and diverse physical phenomena. The local structure and structural coherence across the ferroelectric transition, however, were previously not well understood. Here we reveal the evolution of the local structure with temperature in YMnO$_3$ using neutron total scattering techniques, and interpret them with the help of first-principles calculations. The results show that, at room temperature, the local and average structures are consistent with the established ferroelectric $P6_3cm$ symmetry. On heating, both local and average structural analyses show striking anomalies from $\sim 800$ K up to the Curie temperature consistent with increasing fluctuations of the order parameter angle. These fluctuations result in an unusual local symmetry lowering into a \textit{continuum of structures} on heating. This local symmetry breaking persists into the high-symmetry non-polar phase, constituting an unconventional type of order-disorder transition.
\end{abstract}

\maketitle

\section{INTRODUCTION}\label{sec:introduction}
The multiferroic hexagonal manganites, h-$R$MnO$_3$ ($R$= Dy-Lu, In, Y or Sc) are improper ferroelectrics where the polarization emerges as a secondary effect due to an improper coupling to the primary distortion mode. This results in unusual ferroelectric domain structures in which topological protection of the domain wall intersections causes fundamentally and technologically interesting physical properties ranging from early universe analogues \cite{griffin_scaling_2012,Lin:2014hza, meier_global_2017} to nanoscale conducting channels \cite{fujimura_epitaxially_1996, kumagai_observation_2012,lottermoser_symmetry_2002, choi_insulating_2010, jungk_electrostatic_2010, meier_anisotropic_2012, geng_collective_2012, mundy_functional_2017}. In spite of multiple studies, the evolution of the polarization with temperature has not been explained on a microscopic level. In particular, studies based on powder neutron \cite{gibbs_high-temperature_2011, katsufuji_crystal_2002} and X-ray \citep{katsufuji_crystal_2002, kim_yo_2009,nenert_experimental_2007,tyson_measurements_2011,selbach_crystal_2012, jeong_high-temperature_2007} diffraction show good agreement with a polar model describing the average structure at low and intermediate temperatures, while structural anomalies have been reported between 800 K and the Curie temperature, $T$\sub{C}. This has led to a range of reported values for $T$\sub{C}, and proposals of two distinct structural phase transitions \cite{gibbs_high-temperature_2011, katsufuji_crystal_2002, kim_yo_2009, nenert_experimental_2007, tyson_measurements_2011}, although it is now understood that there is in fact only one phase transition at $T$\sub{C} with the polarization slowly emerging as a secondary effect \cite{fennie_ferroelectric_2005, artyukhin_landau_2013, thomson_elastic_2014, cano_hidden_2014, lilienblum_ferroelectricity_2015}.

\begin{figure*}
\centering
\includegraphics[width=1\textwidth]{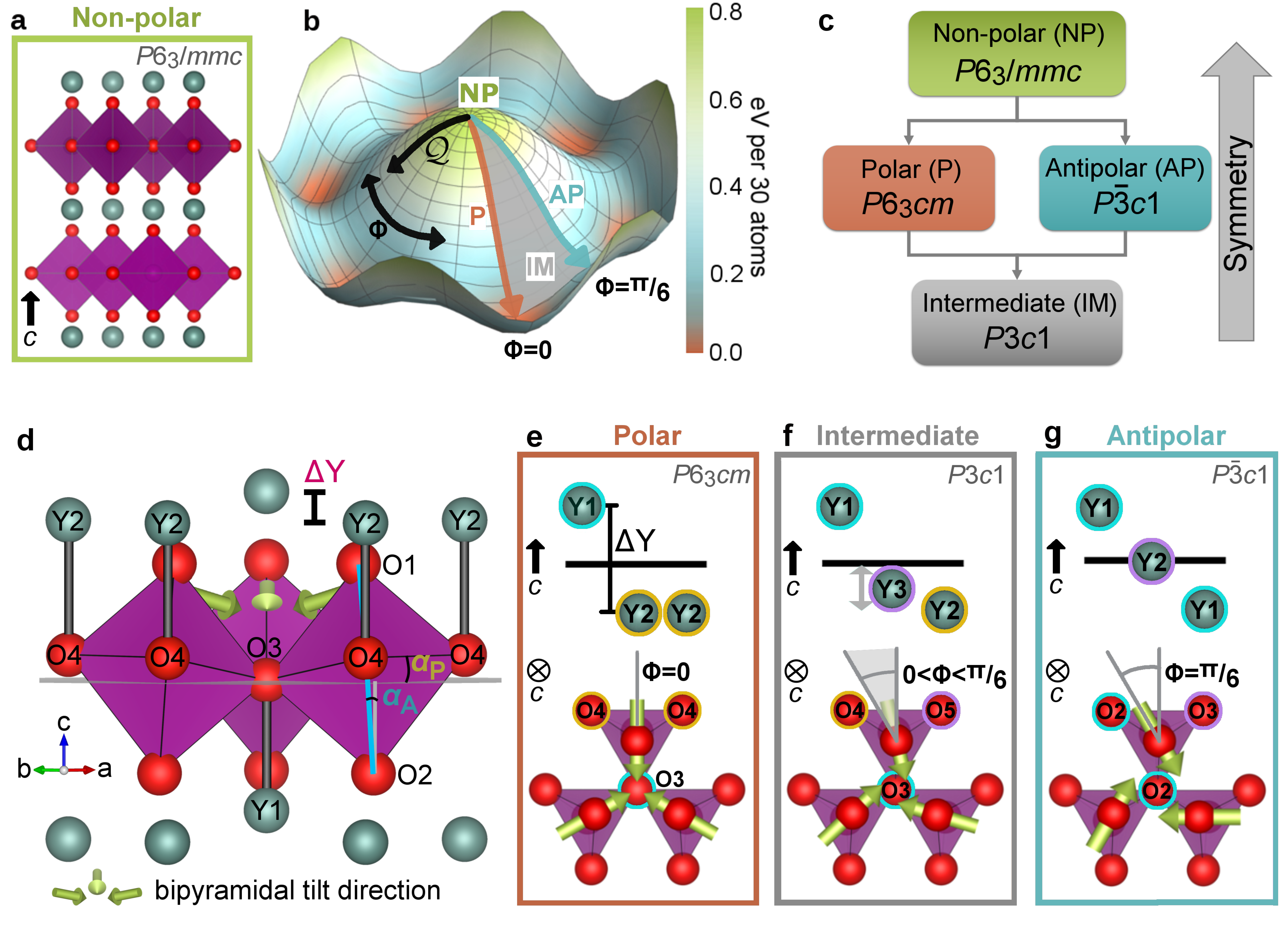}
\caption{Structures of YMnO$_3$ phases. (a) The high-symmetry non-polar (NP) structure with purple MnO$_5$ bipyramids and turquoise Y cations. Oxygens are shown in red. (b) The Landau free energy of the hexagonal manganites as a function of the two-component order parameter ($\mathcal Q$, $\Phi$) resembling a Mexican hat, with the non-polar structure (NP) at the top. In the brim of the hat the polar (P), antipolar (AP) and intermediate (IM) subgroup structures occur at the minima, maxima and intermediate regions, respectively. (c) Group-subgroup relationship between the high-symmetry non-polar structure and the subgroup structures found in the brim of the Mexican hat energy landscape. (d) The order parameter observables for the polar subgroup are the bipyramidal tilt amplitude (angles $\alpha$\sub{A} and $\alpha$\sub{P}) and corrugation of Y cations ($\Delta$Y). $\Delta$Y denotes the distance in the $c$ direction between Y1 and Y2. Green arrows indicate the directions of the bipyramidal tilts. The angle between the O1-O2 line and the $c$ axis defines the apical tilt, $\alpha$\sub{A}, and is a robust measure of the order parameter amplitude $\mathcal Q$ irrespective of the value of $\Phi$. The plane through the three in-plane oxygens (one O3 and two O4) relative to the $ab$ plane defines the planar tilt angle, $\alpha$\sub{P}, and is related to both the order parameter amplitude $\mathcal Q$ and angle $\Phi$. (e)-(g), Structures of the subgroups at different order parameter angles $\Phi$: Y off-centring pattern (top) and bipyramidal tilting directions, indicated by green arrows (bottom). Atomic sites for the three subgroups are labeled and coded with colored circles to emphasize which positions are symmetry related in each phase. Atom positions with the same color markings are aligned along the $c$ axis and have the same multiplicity. A detailed overview of the atomic positions for the space groups is given in Figure S1 of the Supplementary Material \cite{SI_2018}.}
\label{fig:hat}
\end{figure*}

The high-symmetry non-polar average structure of the prototypical hexagonal manganite h-YMnO$_3$ above $T$\sub{C} displays $P6_3/mmc$ symmetry. This structure features corner-sharing MnO$_5$ trigonal bipyramids separated by layers of Y\super{3+} ions, each coordinated by eight oxygens (Fig.~\ref{fig:hat}a). The unit cell triples across $T$\sub{C} at around 1250 K, as a non-centrosymmetric, but zero-polarization, zone-boundary $K_3$ mode condenses \cite{van_aken_origin_2004}. This $K_3$ distortion consists of $z$-axis displacements of the Y atoms, and a corresponding tilt of the trigonal bipyramids. This is described by a two-component order parameter ($\mathcal Q$, $\Phi$) with amplitude, $\mathcal Q$, and angle, $\Phi$ \citep{artyukhin_landau_2013}. $\mathcal Q$ is related to the \textit{magnitude} of bipyramidal tilting and Y displacements, and $\Phi$ is related to the \textit{direction} of the bipyramidal tilting and the displacement \textit{pattern} of the Y.

The resulting energy landscape resembles a Mexican hat (Fig.~\ref{fig:hat}b) with three lower-symmetry space groups below the peak. The group-subgroup relationships are illustrated in Fig.~\ref{fig:hat}c. The subgroups are characterised by the angle of the order parameter $\Phi$. A general angle leads to $P3c1$ symmetry (IM, Fig.~\ref{fig:hat}e), but special values of $\Phi$ lead to higher symmetries. For $\Phi$=$n\frac{\pi}{3}$, with $n$=0,1,..,5, the system is polar with space group symmetry $P6_3cm$ (P, Fig.~\ref{fig:hat}d-e). For $\Phi$=$\frac{\pi}{6}(2n+1)$, the system becomes antipolar with space group symmetry $P\bar{3}c1$ (AP, Fig.~\ref{fig:hat}g). 

Landau theory analysis of the $K_3$ mode following Artyukhin \textit{et al.}~\citep{artyukhin_landau_2013}, in combination with first principles calculations of hexagonal YMnO$_3$, suggests an insignificantly small energy difference between the polar and antipolar symmetry when only the $K_3$ mode is included. Stabilization of the polar state occurs only when the $K_3$ mode couples to a polar $\Gamma_2^-$ mode that causes a shift of the Y atoms towards the Mn-O layer. Due to this improper coupling, nearly all $\Phi$ angles result in a net polarization that reaches its maximum in the polar structure (P), but vanishes for the antipolar structure (AP). The polar structure (P) becomes favoured by $\sim$100 meV per unit cell in comparison with the antipolar (AP) structure, and by more than 600 meV per unit cell in comparison with the non-polar structure (NP) at 0K~\citep{artyukhin_landau_2013}. The exact values of the relative energies are highly temperature dependent. This improper mechanism leads to a polarization of around $\sim$6 $\mu$C cm$^{-2}$ at room temperature. The resulting $Z_6$ symmetry of the configuration space causes the unusual six-fold ferroelectric domain patterns characteristic of h-YMnO$_3$ \cite{safrankova1967}.

This established model \cite{artyukhin_landau_2013} of the Mexican hat Landau free energy (Fig.~\ref{fig:hat}b) describes the average symmetry evolution of the system reduced to the degrees of freedom given by the order parameter, but does not address the underlying microscopics. In particular, whether the transition mechanism is closer to the \textit{displacive} limit (order parameter ($\mathcal Q$, $\Phi$) goes to zero both locally and on average at $T$\sub{C}) or \textit{order-disorder} limit (local order parameter conserved) is not known. 

Here we provide such a local-structure description of the atomic structure of h-YMnO$_3$ from ambient temperature to 1273 K, across the ferroelectric transition $T$\sub{C}. We combine pair distribution function (PDF) analysis of neutron total scattering data with conventional Rietveld refinement to probe structural coherence, and to distinguish short range from average long-range order. 

This analysis reveals a surprising and unconventional behavior of the local structure as a function of temperature that cannot be explained either by a conventional order-disorder or by a displacive transition picture. In a conventional order-disorder transition, the low-temperature distortions persist above the transition in the local structure, but are not evident in the long-range ordered structure due to averaging over variants of the distorted structure \cite{kwei_pair-distribution_1995, keen_crystallography_2015,senn_emergence_2016}. 

Our PDF results show that below $\sim$800~K the average and local structure evolve consistently, with smoothly decreasing distortions. However, between $\sim$800~K and $T$\sub{C}, the average and local structures diverge from each other progressively. The fit of the local structure to the polar model  gradually deteriorates, whilst fits to the antipolar structure gradually improve, consistent with increasing fluctuations of the order parameter angle, $\Phi$, on heating.

At $T$\sub{C}, where long-range cell tripling disappears in the average structure, the fits of the polar and antipolar models become equivalent, though  neither model fits ideally. The transition at $T$\sub{C} is therefore an order-disorder transition in the sense that the average structure transitions to the undistorted high-symmetry non-polar structure whilst the structural distortions persist in the local structure. In other words, the amplitude $\mathcal Q$ of the order parameter never goes to zero. However, it is unconventional in the sense that the disordering does not occur only between the local polar variants.

These two key discoveries -- the unconventional nature of the order-disorder transition and the extensive temperature region of symmetry-lowering fluctuations below $T$\sub{C} where the material accesses a wide range of structures, intermediate between the polar and antipolar subgroups, at the local scale -- reconcile previous literature inconsistencies. 
\\

\section{STRUCTURAL DESCRIPTION OF THE ORDER PARAMETER}
The order parameter ($\mathcal Q$, $\Phi$) is related to three observable atomic displacements (Fig.~\ref{fig:hat}d.): the first, $\alpha$\sub{A}, is the bipyramidal tilt angle  calculated from the line connecting the apical oxygens (O1, O2) and the $c$ axis, the second, $\alpha$\sub{P}, is the bipyramidal tilt  of the plane through the three planar oxygens (O3, O4) relative to the $ab$ plane, and the third is the out-of-plane off-centering/corrugation of the yttrium ions, $\Delta$Y $=c$ $(z_{\text{Y}1}-z_{\text{Y}2})$. The latter two are strictly defined only for the polar ground state symmetry (Fig.~\ref{fig:hat}d). Shifting the order parameter angle $\Phi$ away from the polar symmetry and towards the antipolar or intermediate space groups, breaks the symmetry of the Y2 and O4 sites such that the corresponding order parameter observables $\alpha$\sub{P} and $\Delta$Y have to be calculated by including the additional Wyckoff sites (See Figure S1 in the Supplementary Material \cite{SI_2018}).

\begin{figure}
\includegraphics[width=0.5\textwidth]{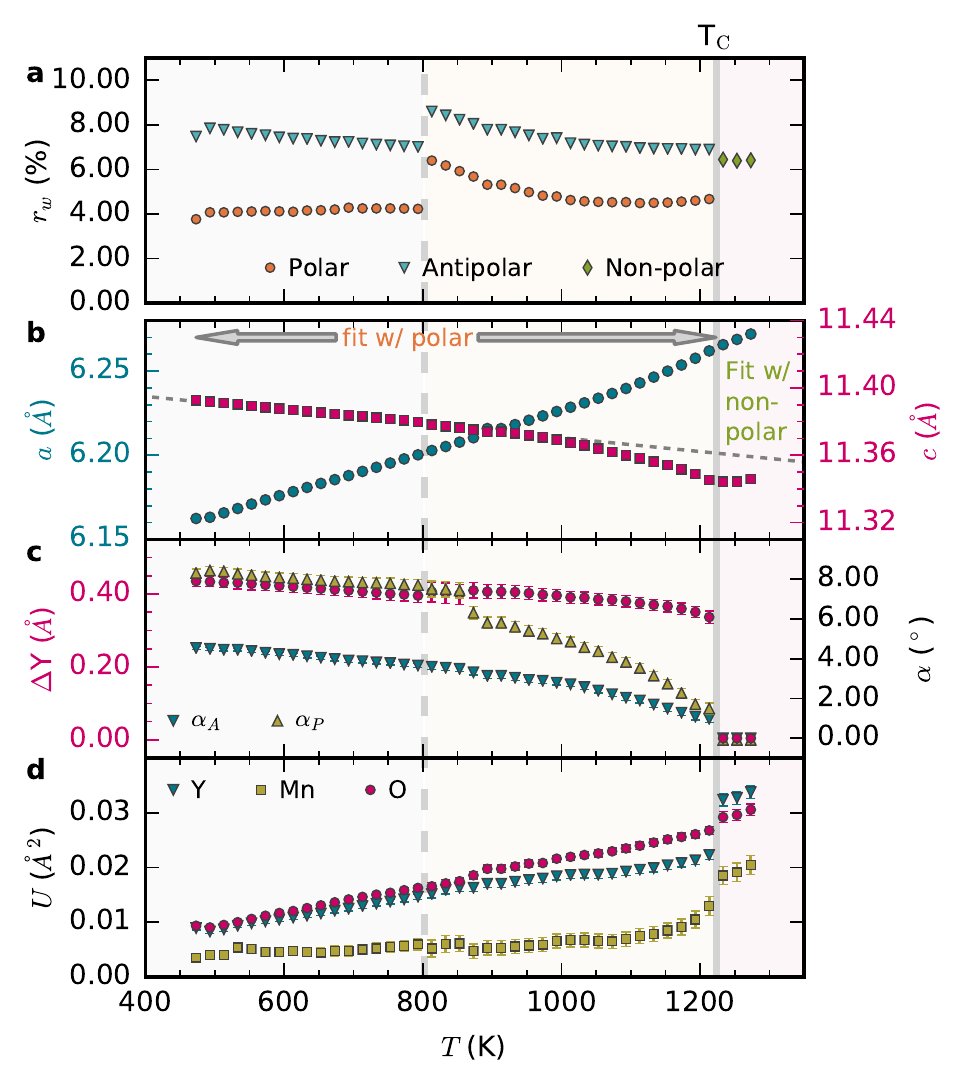}
\caption{Average structure refinements of YMnO$_3$. (a) Fitting residuals $r_w$ of the high-symmetry non-polar (NP) and low-symmetry polar (P) and antipolar (AP) models to $I$($Q$), where $Q$ is the reciprocal space vector. The grey vertical line at 1223 K indicates the Curie temperature $T$\sub{C} above which the high-symmetry non-polar structure can be inferred from the Bragg reflections. The gray vertical line at 800 K separates low- and intermediate-temperature regions. (b) Lattice parameters $a$ and $c$, (c) Y corrugation, $\Delta$Y, apical tilt angle, $\alpha$\sub{A}, and planar tilt angle, $\alpha$\sub{P}, and (d) isotropic atomic displacement parameters, $U$, all obtained from fitting reciprocal space neutron time-of-flight data to the polar ground state model (P) below $T$\sub{C} and to the high-symmetry non-polar structure (NP) above $T$\sub{C} (Representative Rietveld fits of neutron time-of-flight data are shown in Figure S3 in the Supplementary Material \cite{SI_2018}.)}
\label{fig:average}
\end{figure}

\section{RESULTS}
\subsection{Average structure}\label{sec:average structure}
We first consider the average structure behavior based on qualitative and quantitative assessment of our neutron powder diffraction data. On the qualitative side, our reciprocal space data show that the (102) and (202) super-reflections of the polar structure disappear above 1223 K, corresponding to the ferroelectric Curie temperature $T$\sub{C} (See Figure S2 in the Supplementary Material \cite{SI_2018}), in line with previous reports \citep{gibbs_high-temperature_2011, nenert_experimental_2007, jeong_high-temperature_2007, lilienblum_ferroelectricity_2015}.

Quantitative analysis of the average structure of $h$-YMnO$_3$ was carried out by Rietveld refinements  of the measured diffraction patterns. We fitted the high-symmetry non-polar model (NP) at the top of the Mexican hat to the data above $T$\sub{C} = 1223 K. Below $T$\sub{C} and all the way down to 473 K we fitted the polar ground state model (P). We also fitted the antipolar, but distorted, structure (AP) at the local maxima in the brim of the Mexican hat over the same temperature range.  The models are described and visualized in Fig.~\ref{fig:hat}e and g. The results of the fits are summarized in Fig.~\ref{fig:average}, and representative fits are shown in Figure S3 of the Supplementary Material \cite{SI_2018}.

The models gave adequate fits to the data with weighted profile agreement factors ($r_w$) below 10~\%. The polar model gave better agreement than the antipolar model for all temperatures below $T$\sub{C} (Fig.~\ref{fig:average}a). Despite having fewer variables, the non-polar model (NP) gave a comparable $r_w$ to that of the polar model (P) above $T$\sub{C}. The most surprising observation is a discontinuous jump in $r_w$ extracted from the polar (P) and antipolar (AP) models at 800 K. The lattice parameters (Fig.~\ref{fig:average}b) vary smoothly with temperature, also through $T$\sub{C}. Above $T$\sub{C}, lattice parameter $a$ of the non-polar model is multiplied by $\sqrt{3}$ for direct comparison with the tripled unit cell of the polar model.

There is no discontinuity of the lattice parameters at 800 K that might explain the change in $r_w$ there. However, there is a smooth drop-off from linear temperature dependence of the $c$-axis parameter with an onset at around 800 K. This is highlighted in the figure by the gray dashed-line extrapolated from the low-temperature linear behavior. Such a drop-off has been noted previously~\cite{gibbs_high-temperature_2011}, as has the extended region of zero thermal expansion of the $c$-axis above $T$\sub{C}.

We now turn to the three order-parameter observables
$\alpha$\sub{A}, $\alpha$\sub{P} and $\Delta$Y, whose temperature dependencies are shown in Fig.~\ref{fig:average}c. All these parameters have by definition a value of 0 in the high-symmetry non-polar structure and are thus expected to decrease smoothly to zero on heating to $T$\sub{C}, as they are all direct observables of the order parameter amplitude $\mathcal Q$ (Fig.~\ref{fig:hat}). The refinements of the $\alpha$-parameters to the polar model smoothly decrease with increasing temperature, becoming close to zero at $T$\sub{C}, consistent with a continuous transformation. On the other hand, the corrugation parameter, $\Delta$Y, keeps a large value right up to $T$\sub{C}, which would be expected for a discontinuous transition. Again, there are no discontinuous anomalies at 800 K. However, around this temperature the $\alpha$-parameters start to decrease faster than linearly, in agreement with previous studies~\cite{gibbs_high-temperature_2011}.

The vanishing polyhedral tilting and persistent $\Delta$Y, could indicate a scenario in which the structure has off-centered Y ions combined with untilted Mn-O$_5$ polyhedra, which would give very long out-of-plane Y-O distances. We thus performed density functional calculations to check if this scenario is plausible, and found that it is highly unfavourable (See Figure S4 in the Supplementary Material \cite{SI_2018}). Hence, we conclude that the structural behavior obtained from the reciprocal space refinements is physically unfeasible at the local scale, motivating the use of a local structure-sensitive method. 

We note that the Y and O isotropic atomic displacement parameters (ADPs), $U$, shown in Fig.~\ref{fig:average}d are anomalously large compared to the Mn $U$, which is non-linear upon approaching $T$\sub{C}. There is also a discontinuous jump in the ADPs on moving into the high-symmetry non-polar structure, suggesting that broken local symmetry persists above $T$\sub{C}. \cite{tucker_dynamic_2001,egami_underneath_2003,chupas_probing_2004,qiu_orbital_2005}. We therefore turn to a study of the local structure in this material through PDF analysis to further explore this.\\

\subsection{Local structure}\label{sec:local structure}
In order to investigate the local structure, we performed PDF analysis on the same neutron scattering data as analyzed in the previous section. Representative PDFs at different temperatures are shown in Fig.~\ref{fig:pdf}a plotted over a wide range of $r$ (detailed plot Figure S5 in the Supplementary Material \cite{SI_2018}). 

PDF utilizes both Bragg and diffuse scattering information, revealing time-averaged snapshots of the atomic structure on multiple length scales. In contrast to a conventional crystallographic approach that seeks the highest possible symmetry model consistent with the Bragg data component only, the PDF analysis explores whether such symmetry is broken on a nanometer length scale, and if yes, how. The peak positions correspond to interatomic distances, while the widths of the peaks indicate distributions of interatomic distances due to thermal motion, and also lower symmetry. 

Characteristic for neutron PDFs of materials containing manganese (and other negative neutron scattering length materials) is that peaks corresponding to Mn-non-Mn pairs are negative. This is most clearly seen for the Mn-O nearest neighbor bond-length at around 2~\AA. This peak is a doublet as is clearly seen in the 298~K data in Fig.~\ref{fig:pdf}b. On heating the doublet broadens into an unstructured peak by 1273 K (Fig.~\ref{fig:pdf}c and d). The increased broadening of the PDF with increasing temperature is clearly evident in all panels of this figure.

\begin{figure}
\includegraphics[width=0.5\textwidth]{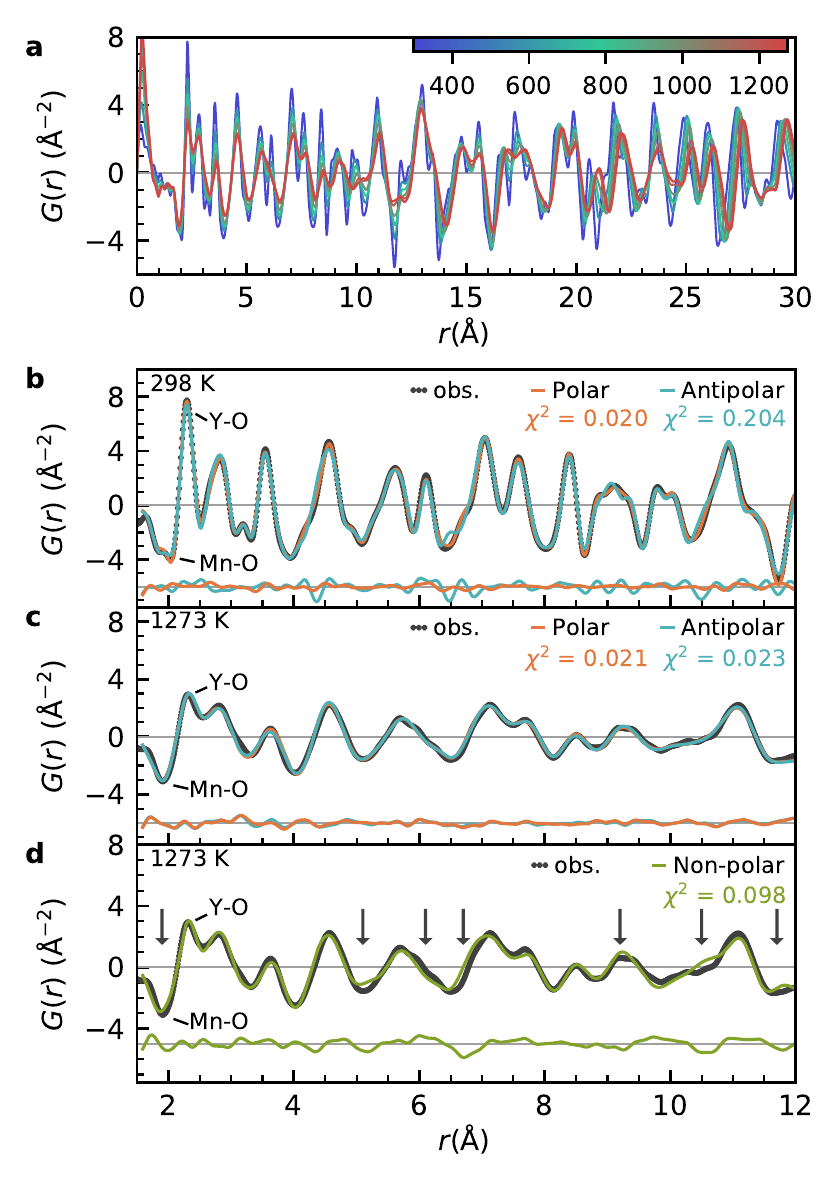}
\caption{PDF of the local structure of YMnO$_3$. (a) Temperature series of the measured $G(r)$ between room temperature and 1273 K in the range $r$ = 0-30 \AA. The horizontal colour bar indicates temperature. (b) The $G(r)$ at 298 K fitted between $r$ = 1.6-12 \AA~ to the low-symmetry polar (P) and antipolar (AP) models. Difference curves are plotted below and overall fit residuals $\chi^2$ are stated close to the legend. The polar model fits the low-temperature data well, while the antipolar model fits worse. (c) Fit of the $G$($r$) between $r$ = 1.6-12 \AA~ at 1273 K to the low-symmetry polar (P) and antipolar (AP) models. (d) Measured $G$($r$) between $r$ = 1.6-12 \AA~at 1273 K fitted by the high-symmetry non-polar (NP) model. Arrows indicate regions of particularly bad fits. The fit residuals below show that this non-polar model does not fit the high-temperature data well at the local scale. Fits of the PDF data for additional temperatures are given in Figure S7 and S8 of the Supplementary Material \cite{SI_2018}.}
\label{fig:pdf}
\end{figure}

We study the structural transition at $T$\sub{C} in a model-independent way by comparing the measured PDF just below and above this transition, as shown in Figure S6 of the Supplementary Material. \cite{SI_2018}. Normal behavior for the PDF is for peaks to broaden on increasing temperature but to sharpen on going to a higher symmetry structure. It is evident in Ref. \cite{SI_2018} that there is no sharpening of the PDF peaks at $T$\sub{C}; indeed careful inspection indicates a thermal broadening in the 1253 K dataset compared to 1193~K that would be expected for a 60 K change in temperature in the absence of a local symmetry raising transition. This clearly shows that the local structure remains distorted in the phase above $T$\sub{C}, despite the crystallographic structure transitioning to the high-symmetry non-polar phase on average.

Quantitative information about the local structure can be extracted from PDF data by fitting to structural models. Because the local structure may display lower symmetry than the average structure, for example, if there are symmetry-broken, but orientationally disordered, domains present \cite{kwei_pair-distribution_1995}, this is often handled by fitting lower-than-crystallographic symmetry models to the PDF data. We fitted the data with models corresponding to the three special space groups found in the Mexican hat energy landscape: the polar (P), antipolar (AP) and high-symmetry non-polar (NP) models over an $r$-range of 12 \AA~corresponding to 1-2 unit cells. 

The representative fits in Fig.~\ref{fig:pdf} show the polar (P) and antipolar (AP) models fit to the room temperature data in panel b, and the high-symmetry non-polar (NP) model fit to the PDF measured above $T$\sub{C} in panel d. Panel c shows the same high-temperature dataset fit with the polar (P) and antipolar (AP) models. We note that, in agreement with the qualitative analysis, the distorted polar model gives much better agreement with the 1273 K dataset than the high-symmetry non-polar model, as evidenced by a lower $\chi^2$ and clear appearance of misfit regions in the difference curve plotted below. The antipolar structure fits significantly worse to the measured PDF at room temperature (Fig.~\ref{fig:pdf}b) but gives comparable fits to those of the polar model at 1273 K. We will return to this observation later.

\begin{figure}
\includegraphics[width=0.5\textwidth]{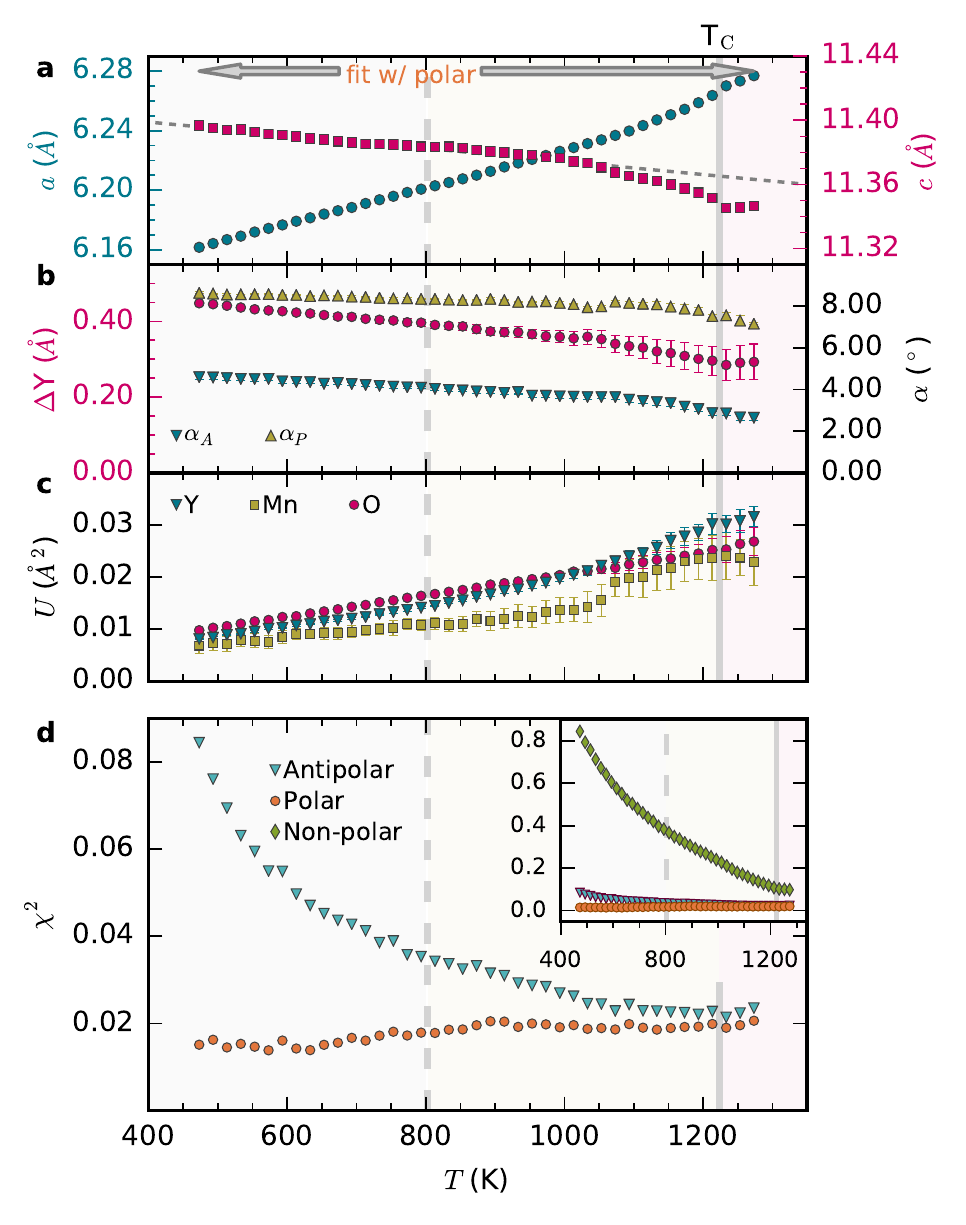}
\caption{Local structure refinements. (a) Refined lattice parameters from fitting $G(r)$ for $r$ = 1.6-22 \AA~with the polar space group model. (b) Y corrugation, $\Delta Y$, planar $\alpha$\sub{P} and apical $\alpha$\sub{A} bipyramidal tilt angles and (c) atomic displacement parameters $U$ from fitting the polar model to the data between $r$ = 1.6-12 \AA~with fixed lattice parameters shown in panel (a) (See Supplementary Methods \cite{SI_2018} for details on calculation of errorbars). (d) Fit residual $\chi^2$ for fitting $G$($r$) between 1.6-12 \AA~for the polar (P) model found at the minima and the antipolar (AP) model found at local maxima in the brim of the Mexican hat potential using fixed lattice parameters found from refinements between 1.6-22 \AA. The grey vertical line at 1223 K shows the Curie temperature $T$\sub{C}, the line at 800 K shows the temperature where refinements become incoherent. Right inset shows fit residuals of the two subgroup models compared to the non-polar (NP) high-symmetry model. A similar plot for the antipolar space group is given in Figure S9 of the Supplementary Material \cite{SI_2018}.}
\label{fig:PDFparameters}
\end{figure}

Next, we consider the lattice parameters and the order parameter observables $\alpha$\sub{A}, $\alpha$\sub{P} and $\Delta$Y, shown in Fig.~\ref{fig:PDFparameters}. The refined lattice parameters are in very good agreement with those obtained from Rietveld refinement, and show the same behavior. However, our local structure refinements show that the order parameter observables do not behave as in the average structure. Whereas $\alpha$\sub{P} and $\alpha$\sub{A} go smoothly to zero at $T$\sub{C} in the average structure, as expected in a purely displacive transition, they both remain finite in the local structure, with $\alpha$\sub{A} decreasing to about half its low-$T$ value and $\alpha$\sub{P} retaining about 80\% of its low-$T$ value. $\Delta$Y behaves close to independently of temperature in the average structure refinements, but decreases in the local structure to about 60 \% of its low-$T$ value. This suggests order-disorder behaviour. 

The error bars of $\Delta$Y become significantly larger than the error bars for $\alpha$\sub{A} and $\alpha$\sub{P} above 1000 K, suggesting that Y displacements become progressively more difficult to fit within the polar model with increasing temperature. We note that, as with the average structure refinements, the atomic displacement parameters and/or their error bars experience an anomalous increase above $\sim$800~K, indicative of increasing inadequacy of this model for describing the local structure as temperature increases. 

Finally, we also compare the fits with the polar model (P) to fits with the antipolar (AP) and high-symmetry (NP) models (Fig.~\ref{fig:PDFparameters}d). The antipolar model (AP) gives distinctly worse agreement than the polar model (P) at room temperature. As temperature increases, the antipolar model fits progressively improve while the polar model fits deteriorate. For well-behaved systems, fit residuals are expected to decrease with heating as thermal fluctuations broaden the data (see for example the behaviour of bulk FCC Ni in Figure S10 of the Supplementary Material \cite{SI_2018}), which is not observed for the polar model. At around 1100 K the two models provide a similar, but not ideal, fit quality, suggesting that neither one of the two provides a complete description of the underlying local structure. Notably, anomalous jumps in $U$ are not fully reconciled by either polar (P) or antipolar (AP) models of the local structure despite reasonably low fit residuals, indicating that the underlying local structure is of greater complexity than that described by these low-symmetry models.

We make three key observations on heating: $\chi^2$ for the polar model (P) increases, $\chi^2$ for the polar (P) and antipolar (AP) models become more similar, and $\chi^2$ for the non-polar model (NP), while decreasing, remains large. These observations lead to the conclusion that more local configurations become represented upon heating. Such a scenario is most likely achieved by increasingly large fluctuations of $\Phi$.

We do not provide fits using the intermediate (IM) model for two reasons. First, it does not represent a unique structure, since it corresponds to a continuum of $\Phi$ values, see Fig.~\ref{fig:hat}f. And second, it has too many degrees of freedom \cite{SI_2018} compared to the number of data observables in the 1-12 \AA~range.

In summary, PDF refinements show that the local structure is neither described by the non-polar model (NP), nor does it take any well-defined distorted structure, as evident from the non-zero order parameter observables and fitting errors Fig.~\ref{fig:PDFparameters}b-d. Instead, we propose that the system has large $\Phi$ order parameter fluctuations, corresponding to different structures in the brim of the Mexican hat.

\section{DISCUSSION}
\begin{figure*}
\includegraphics[width=0.8\textwidth]{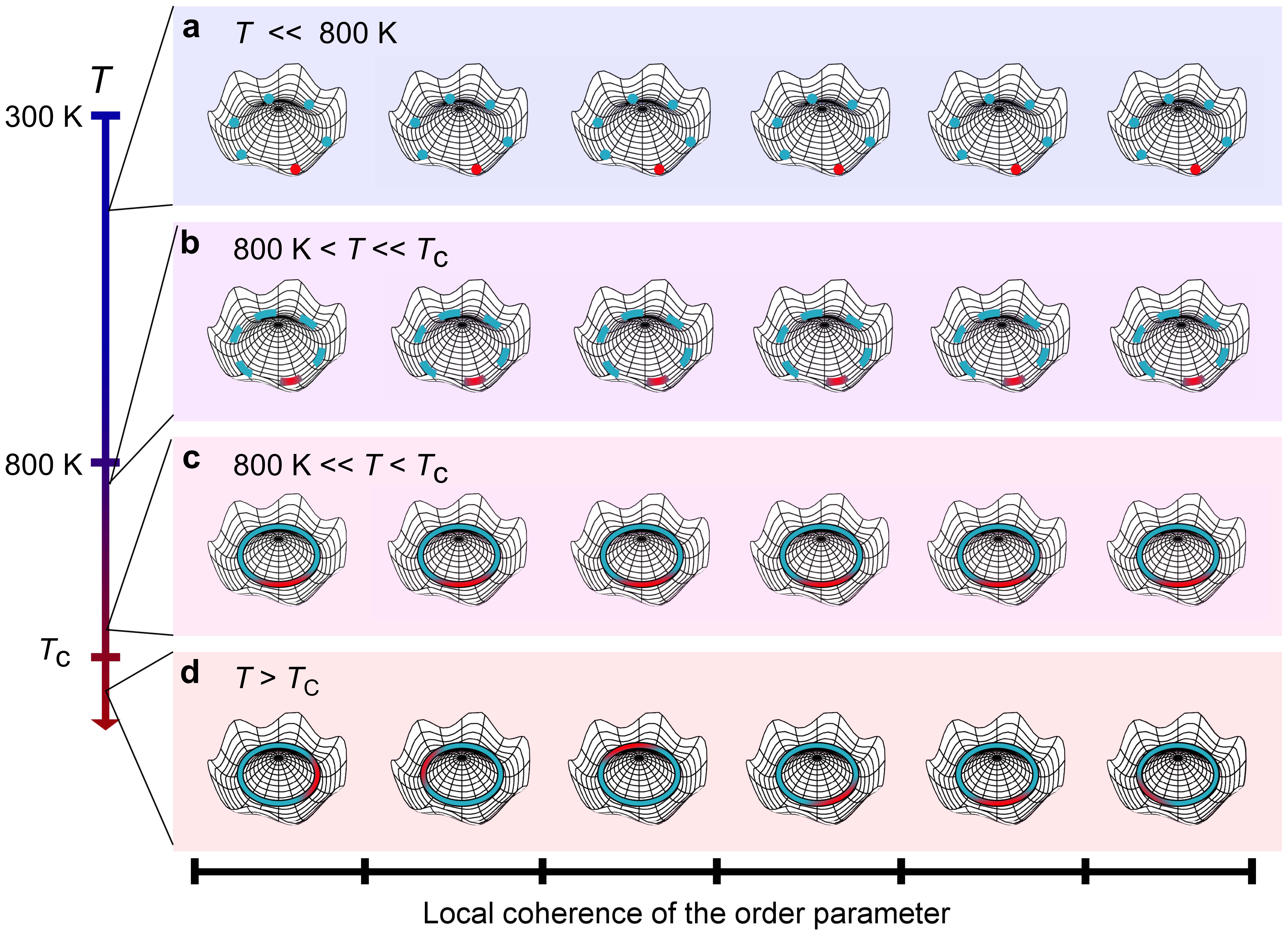}
\caption{Unconventional order-disorder transition of YMnO$_3$. Local value of the order parameter in the different temperature regions (panel a-d) within one domain. The blue lines indicate states accessible to the order parameter and the red colored markings show the distribution of the order parameter. (a) In the low temperature region, the order parameter is firmly fixed within one of the six ground states. (b) Between 800 K and $T$\sub{C} fluctuations smear out the local distribution of the order parameter angle (red), but the average angle within each domain remains constant. (c) Close to $T$\sub{C} the energy landscape becomes flat. (d) At and above $T$\sub{C}, coherence of the order parameter is lost and the system disorders between all possible states in the Mexican hat.}\label{fig:disorder}
\end{figure*}

We have determined the structural evolution with temperature of hexagonal YMnO$_3$ at both local and average length scales. The average structure from reciprocal space Rietveld refinements is captured by the conventional polar (P) model below $T$\sub{C}, and changes to the non-polar (NP) structure across the phase transition, in agreement with previous studies. However, on the local scale the symmetry and the evolution of the structure is very different, and can only be described using the full energy landscape of the two-component order parameter ($\mathcal Q$, $\Phi$), as illustrated in Fig. \ref{fig:disorder}. 

At low temperatures, the local structure corresponds to the polar model (P) of one of the six minima, consistent with the average structure (Fig. \ref{fig:disorder}a).

Upon heating above 800 K, fluctuations of the angle $\Phi$ allow the local structure to access configurations of lower corresponding to the intermediate model (IM) as illustrated Fig. \ref{fig:disorder}b.  This explains the anomalous increase in fit residuals $\chi^2$ for the polar model (P) above 800~K, and the increasingly similar $\chi^2$ for the polar (P) and antipolar (AP) models (Fig \ref{fig:PDFparameters}d). 

Our observations of the progressively equal fit residuals upon heating for the P and AP phases could in principle also be explained by the emergence of higher order terms in the Landau free energy. Such higher order terms have previously been proposed to exist in the hexagonal manganites \cite{cano_hidden_2014}, but they would appear stronger by \textit{increasing} the $\mathcal{Q}$ by lowering the temperature. Since the polar structure (P) shows very good fits at low temperatures and our observations strongly support a \textit{decrease} in $\mathcal{Q}$ upon heating, we do not find this scenario to be likely. We do not find evidence of this hidden order in neither the local nor the average structure.

On approaching $T$\sub{C}, these local fluctuations of $\Phi$ from its mean value become more pronounced. Close to, but below, $T$\sub{C}, the local structure represents a dynamic superposition of states corresponding to many values of $\Phi$ (Fig.~\ref{fig:disorder}c). 

Above $T$\sub{C} (Fig.~\ref{fig:disorder}d), long-range order is lost and the system disorders between a continuum of angles $\Phi$, corresponding to all possible local IM structures with $P3c1$ symmetry. Since the IM structures are polar, the structure remains polar above $T$\sub{C} on the local scale. This special order-disorder transition is non-conventional in the sense that it does not disorder between the six degenerate ground states of the low temperature ferroelectric polar (P) model, but between a $continuum$ of structures with all possible angles $\Phi$ of the order parameter.

This $continuous$ disorder above $T$\sub{C} distinguishes the ferroelectric transition in YMnO$_3$ from conventional order-disorder transitions, e.g. BaTiO$_3$ exhibiting disorder between different, but $discrete$, [111]-oriented ground states \cite{kwei_pair-distribution_1995,senn_emergence_2016}.
The continuous positional degree of freedom in YMnO$_3$ is in some ways reminiscent of $\alpha$-AlF$_3$ \cite{chupas_probing_2004} and cristobalite \cite{tucker_dynamic_2001}, in which the structures disorder with arbitrarily bent bonds between rigid polyhedral units.

The onset of structural fluctuations above 800 K coincides with the temperature of previously reported anomalies in lattice parameters and polarisation \citep{nenert_experimental_2007,gibbs_high-temperature_2011,lilienblum_ferroelectricity_2015,barbour_partial_2016}. Our analysis of PDFs from ambient to above $T$\sub{C} provides a local structure explanation for these anomalies, and also rules out a second macroscopic phase transition. We note that the onset of these fluctuations coincides well with the reported Ginzburg temperature for YMnO$_3$ \cite{meier_global_2017}.

Finally, we note that order-disorder and displacive transitions show different features in their dynamical properties. For a purely displacive transition, the frequency of the soft phonon mode goes to zero at $T$\sub{C}. On the other hand, the soft mode frequency always remains non-zero for a purely order-disorder transition. Instead, one observes the emergence of an additional anharmonic relaxational mode at $\omega=0$, called the central mode \cite{1972shapiro,petzelt_dielectric_1987, Onodera2004,hlinka_coexistence_2008}. Inelastic neutron scattering on YMnO$_3$ performed by Gupta \textit{et al.}~\cite{gupta_spin-phonon_2015} and Bansal \textit{et al.}~\cite{Bansal:2018bf}, showed no critical softening of any phonon branch below $T$\sub{C}, consistent with the order-disorder character inferred from our local structure studies. However, it is not possible to identify a central mode from their data.  

We anticipate that the Mexican energy landscape will lead to unusual dynamics for the central mode. The influence of the central mode on properties like dielectric susceptibility have been studied for proper order-disorder ferroelectrics \cite{Girshberg1997,Weerasinghe2013}, but may be very different in the case of this improper ferroelectric with unconventional disorder. We hope our work motivates future studies in this direction.

\section{CONCLUSIONS}
We have shown that the local structure of YMnO$_3$ across the improper ferroelectric transition at $T$\sub{C} differs strongly from the established average structure symmetry. Fluctuations of the order parameter angle lowers the symmetry upon heating towards $T$\sub{C}, while a finite order parameter amplitude is conserved in the paraelectric phase above $T$\sub{C}. This order-disorder transitions is unconventional as the high-temperature phase disorders between a $continuum$ of structures instead of discrete degenerate minima. The microscopic mechanism of the ferroelectric phase transition of YMnO$_3$ can thus be described by the Mexican energy landscape of the two-component order parameter ($\mathcal Q$, $\Phi$). Our model reconciles previous literature inconsistencies. The onset of order parameter fluctuations coincides with the previously reported second phase transition \cite{nenert_experimental_2007,gibbs_high-temperature_2011,katsufuji_crystal_2002,kim_yo_2009,tyson_measurements_2011}, inferred from thermal expansion anomalies at $\sim 800$ K. Our model is also in agreement with the evolution of the ferroelectric polarization and phonon frequencies across $T$\sub{C} \cite{lilienblum_ferroelectricity_2015,gupta_spin-phonon_2015,Bansal:2018bf,thomson_elastic_2014}.

\section{ACKNOWLEDGEMENTS}
This research used resources at the Spallation Neutron Source, a DOE Office of Science User Facility operated by the Oak Ridge National Laboratory. Computational resources were provided the Euler cluster at ETH, Z\"{u}rich. Financial support from the Research Council of Norway (project no.~231430), NTNU and Advanced Grant (N.A.S.) (no.~291151) from the European Research Council are acknowledged. Work at Brookhaven National Laboratory was supported by US DOE, Office of Science, Office of Basic Energy Sciences (DOE-BES) under contract DE-SC00112704. Matt Tucker and Marshall McDonnell are acknowledged for assisting with the neutron measurements. Matt Tucker is acknowledged for constructing TOPAS input files for reciprocal space Rietveld refinements. J\"{o}rg Neuefeind and Pavol Juhas are acknowledged for assisting manual data reduction and quality testing. Peter Derlet is acknowledged for constructive feedback on the manuscript.\\
\bibliographystyle{apsrev4-1}
\bibliography{ReferenceLibrary}{}

\appendix

\section{Experimental details and data refinement}
Bulk powder of YMnO$_3$ was prepared by firing uniaxially pressed pellets of dried and mixed Y$_2$O$_3$ (>99.99\%, Aldrich) and Mn$_2$O$_3$ (>99\%, Aldrich) twice for 24 h at 1573 K in air with intermediate grinding.

Neutron total scattering was performed at the Nanoscale-Ordered Materials Diffractometer (NOMAD) \cite{neuefeind_nanoscale_2012} at the Spallation Neutron Source at Oak Ridge National Laboratory. Powder sample with mass of $\sim$1.5 g was sealed in a 6 mm diameter vanadium container and measured in an ILL-type vacuum furnace. The NOMAD detectors were calibrated using scattering from diamond standard powder, and Si standard powder was used to obtain the instrument parameter file for Rietveld refinements. The data were collected at room-temperature and between 473-1273 K in steps of 20 K for 60 min.~at each temperature. Measurements of two subsequent temperature cycles between 1113K and 1373 K on the same sample were performed to investigate chemical expansion as a function of oxygen loss, and was found to be insignificant (See Figure S11 of the Supplementary Material \cite{SI_2018}. The structure factor $S$($Q$) was obtained by normalizing the scattering intensity to the scattering from a solid vanadium rod and the background was subtracted using an identical, empty vanadium can. Pair distribution functions (PDF) were obtained by Fourier transform of $S$($Q$) with $Q$\sub{min} = 0.5 \AA\super{-1} and $Q$\sub{max} = 22 \AA\super{-1}.

The average crystal structure over the whole temperature range was determined by Rietveld refinements with the space groups $P6_3/mmc$ and $P6_3cm$ \citep{van_aken_hexagonal_2001}, using TOPAS Academic v.5 \citep{coelho_topas_2004}. PDFs were fitted to the same space groups at different ranges of $r$ using PDFGUI \citep{farrow_pdffit2_2007}. Isotropic atomic displacement factors ($U$) were used \cite{SI_2018}, and more details on calculation of error bars for $\Delta$Y, $\alpha$\sub{P} and $\alpha$\sub{A} are found in the Supplementary Methods. \cite{SI_2018}. We note that error bars for the real space and reciprocal space refinements cannot be quantitatively compared. Details of the PDF method can be found elsewhere \citep{egami_underneath_2003}.

\section{Computational detail}
For our density functional calculations we used the LDA+$U$ approximation \cite{Perdew:1992ee, Liechtenstein:1995ip} as implemented in the abinit PAW planewave code \cite{Gonze:2002br, Gonze:2005em, Torrent:2008jw,Amadon:2008ia}, with a cutoff energy of 30 Hartree and a $k$-point mesh of 6$\times$6$\times$2. A collinear A-type magnetic ordering with a $U$ of 8 eV on the Mn $d$ orbitals was applied. While this magnetic ordering underestimates the bandgap, it does not break the space group symmetry, which is important for this study. 30 atom unit cells with lattice parameters from our PDF refinements were relaxed with respect to atomic positions such that the forces were converged within 1E-6 Ha/Bohr (See Supplementary Methods \cite{SI_2018}).

\end{document}